# Multiclass Permanent Magnets Superstructure for Indoor Localization using Artificial Intelligence

Amir Ivry[1], *Student Member, IEEE*, Elad Fisher[2], Roger Alimi[2], Idan Mosseri[3], and Kanna Nahir[2]

[1]Andrew and Erna Viterbi Faculty of Electrical Engineering, Technion – Israel Institute of Technology, Haifa 3200003, Israel
[2]Technology Division, Soreq NRC, Yavne 81800, Israel
[3]Department of Computer Science, Ben-Gurion University of the Negev, P.O.B. 653, Be'er Sheva, Israel

*Smartphones have become a popular tool for indoor localization and position estimation of users. Existing solutions mainly employ Wi-Fi, RFID, and magnetic sensing techniques to track movements in crowded venues. These are highly sensitive to magnetic clutters and depend on local ambient magnetic fields, which frequently degrades their performance. Also, these techniques often require pre-known mapping surveys of the area, or the presence of active beacons, which are not always available. We embed small-volume and large-moment magnets in pre-known locations and arrange them in specific geometric constellations that create magnetic superstructure patterns of supervised magnetic signatures. These signatures constitute an unambiguous magnetic environment with respect to the moving sensor carrier. The localization algorithm learns the unique patterns of the scattered magnets during training and detects them from the ongoing streaming of data during localization. Our contribution is twofold. First, we deploy passive permanent magnets that do not require a power supply, in contrast to active magnetic transmitters. Second, we perform localization based on smartphone motion rather than on static positioning of the magnetometer. In our previous study, we considered a single superstructure pattern. Here, we present an extended version of that algorithm for multi-superstructure localization, which covers a broader localization area of the user. Experimental results demonstrate localization accuracy of 95% with a mean localization error of less than 1m using artificial intelligence.*

## I. INTRODUCTION

Localization of users indoor has drawn increased interest, with popular methods employ RFID [1] and Wi-Fi [2] techniques. However, magnetic sensors-based approaches recently attracted considerable attention due to their pervasiveness and autonomy [3]. These sensors are mostly utilized in one of two forms: to identify magnetic field characteristics of the infrastructure signatures as locations fingerprints [4], and to employ active magnetic flux transmitters as markers [5].

When using fingerprints, a magnetic map is built by considering magnetic anomalies as landmarks [6]. Thus, the more magnetic inference and anomalies are present, the more enhanced discernability of the magnetic field is achieved. However, pre-known magnetic surveys are required, and simultaneous localization and mapping (SLAM) pose computational burden. Moreover, the fingerprints scheme is highly sensitive to strong electromagnetic interference environments, in which potential localization errors may frequently occur [7]. The second method uses external magnetic field transmitters that produce magnetic signals recorded by the smartphone magnetometer, regardless of existing building magnetic fingerprints [8]. Yet, magnetic flux generation is energy consuming and expensive transmitter units are demanded. Both approaches depend on either active magnetic devices, or suitable mapping surveys of the deployed areas, which renders them sensitive to local magnetic field clutters.

We present a simple, low power, and robust smartphone-based localization in unknown-magnetic-fingerprint indoor locations. We consider a permanent magnets-array that comprises of 3 magnets arranged in a row. We deploy all 6 possible permutations of this array in pre-known locations while preserving the orientation of the magnets. By detecting the unique magnetic signature of each of the 6 patterns using artificial intelligence (AI) algorithms, a passive localization method is obtained.

As the proposed method is landmark-based, we draw comparisons to three competing state-of-the-art approaches for indoor localization with landmark-based magnetism mechanisms, which are covered in [3]. Wang et al. [8] presented UnLoc, established on landmarks matching. MapCraft, introduced by Xiao et al. [9], is rooted on the conditional random-fields method. The IODetector from Li et al. [10] used joint thresholds. For completion, we also refer to localization methods based either on spatial-temporal sequence matching or on fusion with motion [3].

## II. METHODOLOGY

### A. Operating Environment

Consider commercial buildings where many people carry uncontrolled ferrous materials. The magnetic field is measured by the vector magnetometer in the smartphone. Magnetic clutter is present, e.g., stationary magnetic gradients of the construction infrastructure, distorting the natural magnetic flux and causing spoof localizations. The infrastructure may contain large-scale electrically conductive loops acting as sources of magnetic alternating transmissions [11]. Furthermore, moving ferrous objects create interference either from inside the building (people carrying ferrous objects) or from outside (cars, trains). Finally, the smartphone itself generates clutter, both by moving in the environmental field and from its uncontrolled internal electrical currents [11].

### B. Physical Background

Ferromagnetic bodies have the property to deform the ambient magnetic field recorded by the sensor. The signal shape depends on the trajectory of the object, the magnetic moment of the ferromagnetic object, and its position relative to the position of the sensor. The three-axial components of the field $B_x, B_y$ and $B_z$ are given by [11]:

$$\begin{pmatrix} B_x \\ B_y \\ B_z \end{pmatrix} = \frac{\mu_0}{4\pi R^5} \begin{vmatrix} 3x^2 - R^2 & 3xy & 3xz \\ 3yx & 3y^2 - R^2 & 3yz \\ 3zx & 3zy & 3z^2 - R^2 \end{vmatrix} \begin{pmatrix} M_x \\ M_y \\ M_z \end{pmatrix}, (1)$$

where $x = x_{sensor} - x_{object}$ is the x-axis distance between the sensor and object (same for $y$ and $z$), $R$ is the norm distance between the sensor and the object, $M_x$, $M_y$, and $M_z$ are the axial moment components, and $\mu_0$ is magnetic permeability. Usually, dipole approximation is acceptable. Here, the anomaly magnetic flux decreases with inverse proportion to the third power of $R$. Namely, the decay of the magnetic field norm $B$ is:

$$B = \mu_0 \frac{K \cdot M}{R^3} [\mu T], \quad (2)$$

where $\mu_0$ is the magnetic constant, $K$ is a constant with range of 0.1-0.2 and $M$ is the total magnetic moment in Am². Typical values of indoor background clutter (e.g. corridor surroundings) lie around 10-20$\mu T$, easily detectable by smartphone magnetometers with common sensitivity of about 1$\mu T$. To generate patterns with high signal-to-interference-ratio (SIR), an induced field of 40-50$\mu T$ is required. Following Eq. (2), for $R$ between 0.5-1m, a magnet of 125Am² is required. Such devices exist: fixed neodymium magnets, which are powerful mainly due to the tetragonal $Nd_2Fe_{14}B$ crystal structure, having exceptionally high uniaxial magneto-crystalline anisotropy. These magnets are made from magnetic material combining an alloy of neodymium, iron, and boron to form the $Nd_2Fe_{14}B$ structure. To increase the performance of both neodymium and samarium cobalt magnets, traces of additional earth elements such as dysprosium (Dy) and praseodymium (Pr) are added. Then, by considering neodymium magnets with geometric dimensions of $30 \times 30 \times 50$mm and a grade of N35, the magnets produce a Gauss rating of 5,500 over 7,800 times stronger than that produced by the Earth at its magnetic poles. This allows using small but very prominent magnets as passive beacons or markers, producing the required signals. The magnetic field is three-dimensional, and the phone sensor can measure each one of its components.

### C. Multiclass Permanent Magnets Method

We embed small-volume-large-moment permanent magnets in different locations inside the building and arrange them in specific geometric configurations. Practically, we consider 3 neodymium magnets arranged in a row, and embed each of their 6 possible permutations in different locations, without changing their orientation. This results in super-structure patterns of supervised magnetic signatures, constituting unambiguous magnetic environments. The localization algorithm learns these unique patterns during a training stage and detects them from ongoing data streaming during real-time localization, i.e., the testing stage. Localization is based on smartphone motion rather than on static positioning of the magnetometer. Each of the 3 permanent magnets can be approximated by a dipole or a quadrupole, depending on its geometry. These dipoles change the orientation of the flux lines of the Earth magnetic field, creating a unique pattern that can be tuned by ordering the array of magnets in different permutations. A magnetic sensor passing along a path of coded magnets can determine its unique position by matching the pre-learned code. In our previous work [12], we considered a single superstructure, which limits the localization area of the algorithms. In this study, we extend this work by considering 6 distinct patterns, allowing us to cover a broader area in which the user is localized.

### D. Problem Formulation

Let $B_x(n), B_y(n), B_z(n)$ denote the three-axial magnetic signals recorded by the smartphone, where $n \in \mathbb{N}_0$ is a discrete time index. These signals are processed with overlapping time frame of length $N$. We denote $\boldsymbol{f}_i = [\underline{B}_x^i, \underline{B}_y^i, \underline{B}_z^i] \in \mathbb{R}^{1 \times 3N}$ as the extended $i$th time frame that contains concatenation of the $i$th frame of each axis signal. Assume 7 hypotheses for each frame $\boldsymbol{f}_i$, notated $\{\mathcal{H}_0, \ldots, \mathcal{H}_6\}$. The hypothesis $\mathcal{H}_0$ represents frames that do not contain any part of a superstructure magnet. The hypothesis $\mathcal{H}_j$, represent frames that include a part of the superstructure magnet $j$, where $j \in [1,6]$. The goal of this study is to correctly identify the hypothesis of each frame.

### E. AI-based Models

We implement 6 AI-based algorithms with different modeling mechanisms: two variations of support vector machine (SVM) [13], with and without principal components analysis (PCA) [14]. To efficiently perceive non-linear trends, a fully connected deep neural network (DNN) [15] was used with four hidden layers of 400 neurons each. These methods lack feedback information in their mechanisms, which is instructive when dealing with time series modeling. Thus, a recurrent neural network (RNN) [16], a gated recurrent unit (GRU) [17], and long-short term memory (LSTM) [18] networks are also implemented.

## III. EXPERIMENTAL SETUP

### A. Apparatus

Measurements were collected using Xiaomi-mi4 smartphone and the Sensor Kinetics Pro sensor acquisition application, which recorded magnetic sensing information from the BOSCH magnetic field sensor with sensitivity between 0-1600$\mu$T, resolution of 0.3$\mu$T, and power consumption of 0.5mA.

### B. Settings

Experiments were conducted in different locations for the training stage and for the test stage. For the training stage, a shielded environment of negligible magnetic interference was considered, with an average norm field power lower than 0.5$\mu T$. In this setup, each of the 6 superstructures was recorded by passing by it with the smartphone 30 times. For the test stage, 4 indoor environments of different intrinsic magnetic patterns were considered. In each of them, the user recorded 60 minutes of data, comprising 10 passes by each of the embedded magnetic patterns. In both training and test recordings, the user strolled with the smartphone in-pocket in walking paces that differs between 0.8-2m/s and walking distances that ranged between 0.5-1m from the superstructures. The smartphone heading azimuth, i.e., its orientation, ranged between $0°$ and $360°$, while elevation remained unchanged. The magnets were positioned at height of 1.5m and were separated by 3m from one another. The test recordings were performed with an average SIR level of 8dB, where SIR is defined as the power ratio between those frames that contain a superstructure pattern and those which do not. SIR and power calculations were calculated using 50% overlapping time frames of 20ms. More specifically, the constellation of magnets was comprised of one unit of 3 attached magnets, one unit of 2 attached magnets, and one unit of a single magnet. In each of the 6 pattern combinations, these 3 separate units were arranged in a row 3m from each other, and their relative orientations remained unchanged [12].

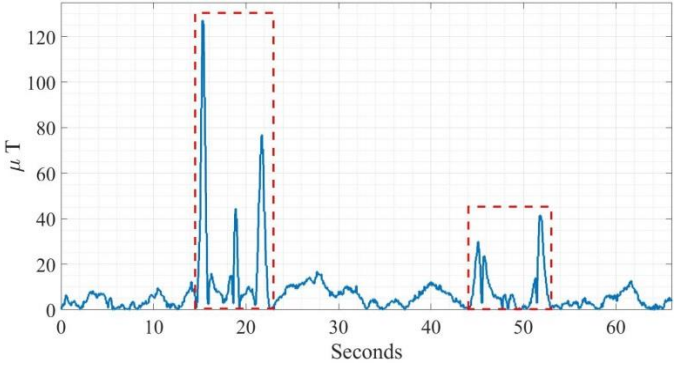

Fig. 1. Norm value of the three-axial magnetic field recorded by the smartphone along 40 seconds, with an average SIR level of 8dB. The confined patterns are recorded as the user crosses the same magnetic superstructure with different orientations between the smartphone and the superstructure. The unconfined recordings capture inherent fingerprints of the indoor environment.

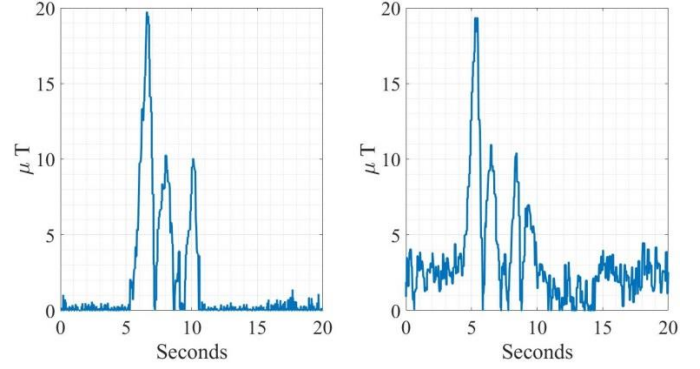

Fig. 2 Norm value of the three-axial magnetic field recorded by the smartphone along 20 seconds. The same superstructure is recorded with the same orientation between the smartphone and the superstructure, in two different locations: in the shielded training environment, and in noisy indoor environment (right) with an average SIR level of 8dB.

The contribution of learning the superstructures in a shielded environment using various device orientations is depicted in Fig. 1 and Fig. 2. Fig. 1 illustrates the effect of the orientation between the smartphone and the magnetic superstructure on the recorded superstructure pattern. Substantial variations are shown between two snippets of the same superstructure, due to device orientation change. Thus, learning recordings of the magnetic superstructures from different device orientations is highly instructive. Also, Fig. 2 shows another magnetic superstructure that was recorded in the training shielded environment on one hand, and in the testing environment on the other hand, with an identical smartphone orientation. The training setup produces a clear pattern that can be learned by the algorithm and later detected in the noisy test environment.

### C. Database, Pre-processing, and Features Extraction

The data recorded from the smartphone contained five signals: the 3-axial magnetic signals, pitch, and roll, notated $B_x, B_y, B_z, P$, and $R$, respectively. The magnetic norm, $B$, was calculated to enhance the prominent energy of instilled magnets:

$$B = \sqrt{B_x^2 + B_y^2 + B_z^2}. \quad (3)$$

We aimed to reduce dependency of measurements on the smartphone orientation. Thus, we generated the horizontal and vertical magnetic components, $B_h$ and $B_v$, respectively [19]:

$$B_v = -\sin(P) \cdot B_x + \sin(R) \cdot B_y + \cos(P) \cdot \cos(R) \cdot B_z, \quad (4)$$

$$B_h = \sqrt{B^2 - B_v^2}. \quad (5)$$

All magnetic signals sense prominent magnetic anomalies around the magnets relative to the intrinsic magnetic environment. While $B$ expectedly captures the explicit target pattern, $B_h$ follows the same behavior while being user-orientation-independent. During pre-processing, rectangular windows of $N = 12.5$s with 80ms shift divided the data into frames, capturing the entire embedded pattern in a single frame, while containing each superstructure multiple times. The training database contains 150k data frames and the test database contains 100k data frames, both are constructed as detailed in Section III(B). We perform feature extraction from the time series measurements to feed the different AI methods with a compact and informative data representation. The extended $i$th time frame is given by:

$$\tilde{f}_i = [B_x^i, B_y^i, B_z^i, B^i, B_h^i, B_v^i] \in \mathbb{R}^{1 \times 6N}. \quad (6)$$

Each one of the 6 signal components in $\tilde{f}_i$ is mapped to 20 identical features with physical and statistical orientations. To facilitate temporal behavior, each feature vector is concatenated to 2 feature vectors affiliated with 2 adjacent past time frames, resulting in $20 \times 6 \times 3 = 360$ features per time frame. Past look of 2 frames enhanced performance without overconsuming computational load. Formally, let $FE(\cdot)$ hold the feature extraction operator and denote $\tilde{x}_i \in \mathbb{R}^{1 \times 360}$ as the feature vector extracted from $\tilde{f}_i$. Then, for $i > 2$:

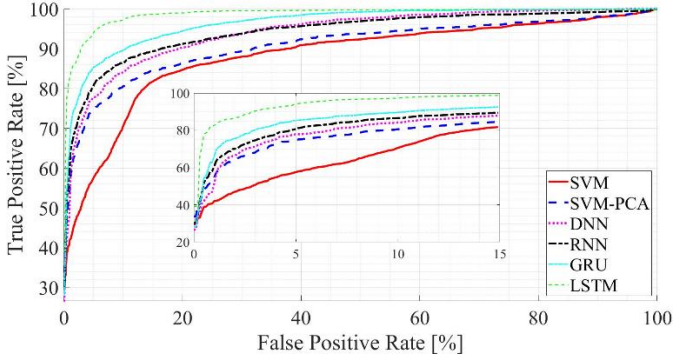

Fig. 3. ROC curve comparison between the performance of all 6 AI-based models for indoor localization. Zoom-in view is attached.

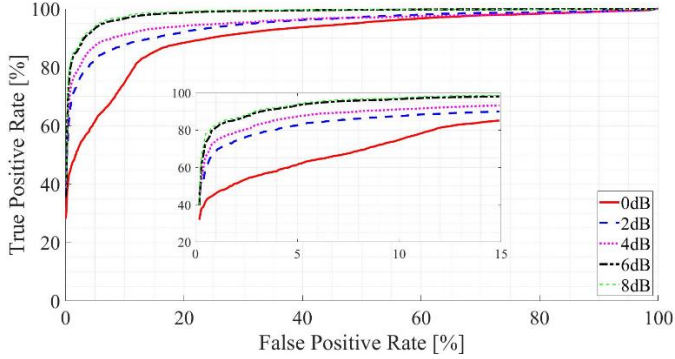

Fig. 4. ROC curve comparison between the performance of the LSTM model in various SIR levels, from 0dB to 8dB. Zoom-in view is attached.

$$\widetilde{x}_i = \begin{bmatrix} FE(B_x^i), FE(B_y^i), FE(B_z^i), FE(B^i), FE(B_h^i), FE(B_v^i), \\ FE(B_x^{i-1}), FE(B_y^{i-1}), FE(B_z^{i-1}), FE(B^{i-1}), FE(B_h^{i-1}), FE(B_v^{i-1}), \\ FE(B_x^{i-2}), FE(B_y^{i-2}), FE(B_z^{i-2}), FE(B^{i-2}), FE(B_h^{i-2}), FE(B_v^{i-2}) \end{bmatrix}. \quad (7)$$

Each feature vector $\widetilde{x}_i$ is linked with a multiclass label $y_i \in \{0, 6\}$, where labels 1 to 6 represent the 6 different superstructures, and 0 represents the absence of a superstructure in that frame.

### D. Training, Validation, and Testing Processes

During training, the AI models learn to identify and distinguish between the 6 superstructures. Also, 80% of the training set is used to train the models, and 20% is used for validation. If the validation accuracy does not increase after 5 epochs, the training process is stopped, and the model is declared optimal. This was done by minimizing the cross-entropy loss function between the AI model output and label, with the Adam optimizer [20]. The batch size used was 64 frames, the learning rate varied between 0.0005 to 0.001, depending on the AI model used. The testing is done by feeding the trained AI models with the test set features. The output of each test frame is compared against its ground truth label, and performance measures are derived.

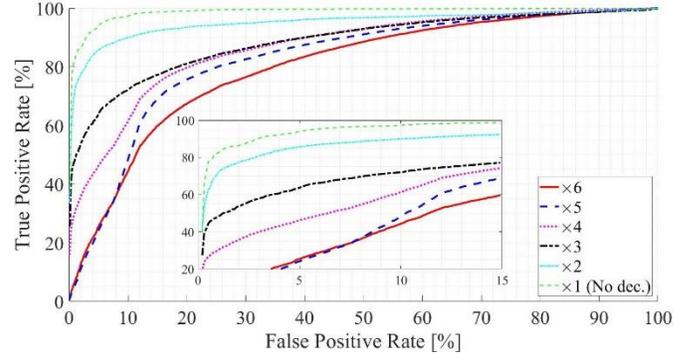

Fig. 5. ROC curve comparison between the performance of the LSTM model in degraded conditions of decimated smartphone resolution. The most degraded scenario inspects sample frequency that is 6 times lower than the original maximum rate. Zoom-in view is attached.

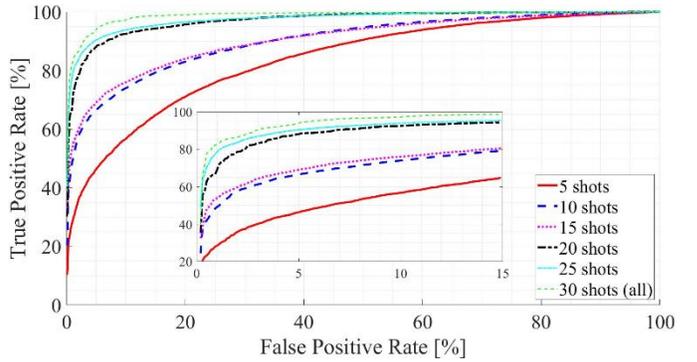

Fig. 6. ROC curve comparison between the performance of the LSTM model in various few-shot learning scenarios during training. It ranges between using only 5 shots up to 30 shots from each of the superstructures during training. Zoom-in view is attached.

## IV. RESULTS AND DISCUSSION

### A. AI-based Magnetic Localization

The localization performance of the 6 AI algorithms is shown in Fig. 3 using the receiver operating characteristic (ROC) curve, allowing analysis of the trade-off between true positive rate (TPR) and false-positive rate (FPR) in various system operation points [21]. The long short-term memory (LSTM) network leads in localization performance, followed by the GRU and RNN. Even though they show degraded performance relative to the LSTM due to less advanced temporal modeling, the GRU and RNN architectures allows lower memory consumption and computational load, being more adequate for on-device applications. The least performant methods are the DNN and SVM. We define the localization accuracy as the maximal TPR + true negative rate (TNR) [22]. LSTM shows 95% accuracy, while GRU and RNN reach 92% and 90%, respectively. DNN shows 83%, SVM-PCA method reaches nearly 80%, and SVM produces 73% accuracy. The superiority of the LSTM over competing methods is explained as follows. LSTMs can learn optimal time series context via feedback-based learning for temporal classification, this context being pre-specified and fixed in the DNN and SVM methods. Even though the RNN employs feedback, it suffers short-term memory, preventing it from modeling temporal context between early and late periods. To mitigate this gap, the GRU and LSTM architectures contain gate mechanisms that regulate information flow through the network. While the GRU contains fewer parameters that require less memory and training data, the LSTM embeds an additional memory unit inside its gate allowing enhanced control over information flow [23].

*B. Comparison to Competing Methods*

For performance evaluation, we employ the key metrics of mean localization error (MLE) [3] and localization accuracy [22], which is employed in Section IV(A). These metrics are sensitive to changes in test environments, so the following values are listed for reference only. UnLoc achieved MLE between 1-2m and outperformed MapCraft. Both approaches show lower performance compared to the proposed approach, localizing at maximal error of 1m. The IODetector was able to obtain localization accuracy as high as 82% indoor, which falls short to the 95% presented by the LSTM-based algorithm used in this study. Additional methods rooted on spatial-temporal sequence matching or on fusion with motion [3], also show lower MLE and classification accuracy compared to our study.

*C. Generalization and Robustness*

Additional experiments were conducted using the leading LSTM algorithm to check the generalization and robustness of the proposed localization method. First, localization accuracy with respect to the walking pace of the carrier was analyzed, to verify if our approach is applicable to real-life scenarios where walking paces vary. Walking paces were taken from the ensemble {0.8, 1.2, 1.6, 2} m/s. Performance analysis showed respective average localization accuracies of {94.1, 96, 95.3, 93.5} %. This evaluation exhibits a high robustness of the proposed approach for both slow and fast paces. The optimal walking pace, in terms of maximal accuracy, is expected to vary depending on the smartphone sampling rate, distance between magnets, and carrier walking distance. Localization should also apply in intrinsic environments of low SIR levels. While experiments dealt with 8dB SIR, we synthetically inspected the performance in SIRs ranging from 6dB to 0dB by decreasing the energy of frames that contained magnetic superstructures. Fig. 4 shows these results. 6dB SIR barely affects the performance with average degradation of less than 2.5% in accuracy, while 4dB SIR leads to an accuracy of 86%, and 0dB SIR reduces the accuracy to 80.5%. Another experiment synthetically decimated the resolution of test recordings to simulate CPU load, which cannot always be compensated, e.g., using interpolation, during real-time on-device usage. Results are illustrated in Fig. 5. While training measurements were taken at 120Hz, lower sample frequencies of the test set of 60Hz and 30Hz, lead to 86% and 73.5% accuracy, respectively. It emphasizes the highly generalized modeling of the LSTM scheme that enables it to identify the target signature with missing information derived from impeded data resolution. In practical scenarios, it is beneficial to record fewer shots of the magnetic pattern during training to allow quick and efficient system re-appliance. As showed in Fig. 6, instead of crossing every magnetic superstructure 30 times, 20 training snippets were sufficient for nearly 90% accuracy, and 10 crosses provide accuracy of nearly 80% accuracy. These experiments demonstrate the robustness of the LSTM algorithm to dominant magnetic interference even with lower amounts of training data. Lastly, we note that the achieved performance may be affected by the inherent measurement error of the smartphone, mentioned in Section III(A). Future research will involve a more extensive comparison between recording devices with different resolutions and their effect on localization accuracy.

## V. CONCLUSION

In this study, low-power dynamic indoor localization method combined with AI is presented. Passive permanent magnets are deployed instead of active magnetic transmitters. The localization was performed based on smartphone motion rather than on the static positioning of the magnetometer. Inspired by our latest work that considered a single magnetic pattern, this work is extended to a multi-class problem of 6 different patterns allowing a broader area of user localization. Among various AI schemes, the LSTM network showed leading performance of 95% accuracy. More experiments point to the generalization and robustness of the LSTM algorithm regarding high magnetic interference and lower amounts of training data. These results are very promising regarding applying passive localization of a moving person in indoor environment. For instance, it may be used for locating potential customers passing by a targeted area in a shopping mall or a store aisle. Further refinement of this method will address specific magnets orientations and improved AI models to reach a more enhanced performance.